# Role of dimensional crossover on spin-orbit torque efficiency in magnetic insulator thin films

Qiming Shao[1#], Chi Tang[2#], Guoqiang Yu[1,3*], Aryan Navabi[1], Hao Wu[3], Congli He[1], Junxue Li[2], Pramey Upadhyaya[4], Peng Zhang[1], Seyed Armin Razavi[1], Qing Lin He[1], Yawen Liu[2], Pei Yang[1,5], Se Kwon Kim[4], Cheng Zheng[1], Yizhou Liu[6], Lei Pan[1], Roger Lake[6], Xiufeng Han[3], Yaroslav Tserkovnyak[4], Jing Shi[2], and Kang L. Wang[1,4*]

[1]Department of Electrical and Computer Engineering, University of California, Los Angeles, CA 90095, United States

[2]Department of Physics & Astronomy, University of California, Riverside, CA 92521, United States

[3]Beijing National Laboratory for Condensed Matter Physics, Institute of Physics, Chinese Academy of Sciences, Beijing 100190, China

[4]Department of Physics & Astronomy, University of California, Los Angeles, CA 90095, United States

[5]York-Nanjing Joint Center for Spintronics and Nano Engineering (YNJC), School of Electronics Science and Engineering, Nanjing University, Nanjing 210093, China

[6]Department of Electrical and Computer Engineering, University of California, Riverside, CA 92521, United States

*corresponding authors: guoqiangyu@iphy.ac.cn and wang@ee.ucla.edu

[#]these authors contributed to this work equally.

Magnetic insulators (MIs) attract tremendous interest for spintronic applications due to low Gilbert damping and absence of Ohmic loss. Magnetic order of MIs can be manipulated and even switched by spin-orbit torques (SOTs) generated through spin Hall effect and Rashba-Edelstein effect in heavy metal/MI bilayers. SOTs on MIs are more intriguing than magnetic metals since SOTs cannot be transferred to MIs through direct injection of electron spins. Understanding of SOTs on MIs remains elusive, especially how SOTs scale with the film thickness. Here, we observe the critical role of dimensionality on the SOT efficiency by systematically studying the MI layer thickness dependent SOT efficiency in tungsten/thulium iron garnet (W/TmIG) bilayers. We first show that the TmIG thin film evolves from two-dimensional to three-dimensional magnetic phase transitions as the thickness increases, due to the suppression of long-wavelength thermal fluctuation. Then, we report the significant enhancement of the measured SOT efficiency as the thickness increases. We attribute this effect to the increase of the magnetic moment density in concert with the suppression of thermal fluctuations. At last, we demonstrate the current-induced SOT switching in the W/TmIG bilayers with a TmIG thickness up to 15 nm. The switching current density is comparable with those of heavy metal/ferromagnetic metal cases. Our findings shed light on the understanding of SOTs in MIs, which is important for the future development of ultrathin MI-based low-power spintronics.

**Introduction**

The interplay between heavy metals (HMs) and magnetic insulators (MIs) in heavy metal/magnetic insulator (HM/MI) bilayer systems has attracted tremendous attention from both fundamental research and practical applications [1-4]. First, the HM/MI bilayer benefits from the low Gilbert damping in MIs. Different from magnetic metal cases, MIs only allow spin information to propagate through magnons, instead of itinerant electrons, due to their large electronic bandgaps. The absence of Ohmic loss from the magnetic layer makes HM/MI bilayers more energy efficient than HM/magnetic metal cases.

The second advantage of the HM/MI bilayer is that the spin-orbit coupling in HM or at HM/MI interface allows the efficient generation of spin-orbit torques (SOTs) on the MI layer through spin Hall effect (SHE) or Rashba-Edelstein effect [5-9]. These SOTs enable efficient manipulation of magnetization dynamics in the MI layer. Although the MI layer is electrically insulating, SOT-driven magnetization dynamics of magnetic insulators can be detected through anomalous Hall resistance (AHR) and spin Hall magnetoresistance (SMR) in the HM layer [10-13]. By probing the AHR, current-induced magnetization switching (CIMS) was realized in both Pt/BaFe$_{12}$O$_{19}$[14] and Pt/Tm$_3$Fe$_5$O$_{12}$ (TmIG) bilayers [15,16]. However, the mechanism of SOTs in Pt/MI bilayers remains ambiguous [16]. Moreover, the observed damping-like SOT efficiency ($\xi_{DL}$) in the Pt/TmIG that is responsible for switching is still much lower than those in the Pt/ferromagnetic metals (FMs) [15,17,18]. To understand SOTs and increase the value of $\xi_{DL}$ in HM/MI bilayers, studying SOTs and realizing magnetization switching in a HM/MI bilayer with a large spin Hall angle that is opposite to that of Pt are essential.

In this Article, we study the $\xi_{DL}$ and CIMS in tungsten (W)/TmIG heterostructures with different TmIG layer thicknesses ($t_{TmIG}$). Thickness-dependent damping-like SOT study allows us to understand the interplay between spin current and magnetism in TmIG. Here, β-W is chosen since it is reported to give the largest spin Hall angle among elemental heavy metals and its sign is opposite to that of Pt [19]. When the TmIG film thickness is reduced from 15 nm to 3.2 nm, the effective exchange coupling is strongly reduced due to long-wavelength thermal fluctuation, resulting in a dimensional crossover from three-dimension-like to two-dimension-like magnetic phase transitions. We quantify $\xi_{DL}$ by using second-harmonic Hall measurements [20,21]. The $\xi_{DL}$ increases with the $t_{TmIG}$ in W/TmIG bilayers; this is attributed to the enhanced magnetic moment density due to suppression of thermal fluctuation. We then demonstrate the CIMS in W/TmIG bilayers up to $t_{TmIG}$ = 15 nm; for $t_{TmIG}$ = 15 nm, the switching current density is as low as $8 \times 10^{10}$ A/m$^2$. The estimated current switching efficiency enhances as $t_{TmIG}$ increases, which is consistent with the increase of $\xi_{DL}$ with $t_{TmIG}$. Importantly, the switching direction of our W/TmIG devices is indeed opposite to that of Pt/TmIG case [15]; this contrast confirms the important role of SHE in CIMS of MIs.

**Dimensional crossover of magnetism**

To access SOT and realize CIMS, we prepare high-quality TmIG thin films with different $t_{TmIG}$ and characterize their magnetic properties. These TmIG(111) thin films were grown on substrate Nd$_3$Ga$_5$O$_{12}$(111) by pulsed laser deposition [13]. All TmIG thin films show an atomically flat surface with mean roughness as low as 0.1 nm (Fig. 1a), providing a sharp interface for efficient spin momentum transfer. The large lattice mismatch between the TmIG and the Nd$_3$Ga$_5$O$_{12}$ provides the tensile strain to generate perpendicular magnetic anisotropy in all TmIG thin films. The nature of perpendicular magnetic anisotropy is confirmed using magnetization hysteresis loops of TmIG thin films as a function of an out-of-plane magnetic field (Fig. 1b), from which we can determine saturation magnetization ($M_S$). We observe a strong $t_{TmIG}$ dependence of the $M_S$ at room temperature (Fig. 1c); the $M_S$ reduces significantly from the bulk $M_S$ (110 emu/cc) [22] with decreasing film thickness. Note that the estimated dead layer thickness is less than 1 nm (see Fig. 1c inset), which also suggests a sharp interface between TmIG and substrate [23]. The reduction of the $M_S$ at room temperature is attributed to finite size effect, strong thermal fluctuation and strong surface modification effect in ultrathin magnetic films [24-26]. Following ref. [25], we extract the critical exponents $\beta$ for magnetic phase transitions in these TmIG thin films using temperature dependence of magnetic moment (*M-T*). The *M-T* curves follow the $M = M_0\left(1 - T/T_C\right)^\beta$ (Fig. 1d), where zero-temperature magnetic moment ($M_0$) and Curie temperature ($T_C$) are fitting parameters. The $t_{TmIG}$ dependent $\beta$ is better illustrated using log-log plots as shown in Fig. 1e and the results are summarized in Fig. 1f. We see a clear increase of $\beta$ from 0.16±0.06 to 0.42±0.02 when the $t_{TmIG}$ increases from 3.2 nm to 15 nm. This increase of $\beta$ suggests a dimensional crossover from two-dimension-like to three-dimension-like magnetism since 2D Ising model and 3D Heisenberg model predict $\beta$ to be 0.125 and 0.365, respectively[26,27]. The dimensional crossover happens at around 6 nm, which is one order of magnitude larger than the typical transition thickness around 1 nm for magnetic metals [25-27]. In the following sections, we point out that the reduction of $M_S$ due to dimensional crossover has a major influence on the magnitude of the SOT and switching efficiency, which has been neglected in the previous experiments.

**SOT measurement**

To perform resistance, SOT and CIMS measurements, we fabricate W(5nm)/TmIG($t_{TmIG}$) thin films into Hall bar devices (Fig. 2a). The AHR in the W/TmIG is accurately determined by the sharp anomalous Hall hysteresis at low fields (Fig. 2b). The transverse planar Hall resistance (PHR) accompanying the longitudinal SMR is measured by rotating the magnetization in the *xy*-plane (Fig. 2c). The observation of sizeable AHR and PHR (SMR) indicates that there is a significant spin current being transmitted across the W/TmIG interface or a sizable spin mixing conductance [11] (see supplementary information).

We quantify $\xi_{DL}$ by using the second-harmonic analysis of both AHR and PHR ($R_{AHE}$ and $R_{PHE}$) [20,21]. The second-harmonic Hall resistance ($R_H^{2\omega}$) in a single domain subjected to an in-plane magnetic field can be written as [21,28]

$$R_H^{2\omega} = R_{FL}^{2\omega} \cos 2\varphi \sin \varphi + R_{DL}^{2\omega} \sin \varphi = R_{PHE} \frac{H_{FL}}{|H_{ext}|} \cos 2\varphi \sin \varphi + \frac{R_{AHE}}{2} \frac{H_{DL}}{|H_{ext}|-H_K} \sin \varphi, \quad (1)$$

where $H_K$ and $H_{ext}$ are perpendicular magnetic anisotropy effective field and in-plane external field. In Eq. (1), $R_{FL}^{2\omega}$ and $R_{DL}^{2\omega}$ are the peak values of $\cos 2\varphi \sin \varphi$ and $\sin \varphi$ components in $R_H^{2\omega}$, which are field-like SOT and damping-like SOT contributions, respectively. $H_{FL}$ and $H_{DL}$ are the current-induced field-like and damping-like effective fields, respectively. For example, when the $H_{ext}$ = 2500 Oe, we observe significant contributions from both damping-like and field-like SOTs, as reflected by the $\cos 2\varphi \sin \varphi$ and $\sin \varphi$ angle dependencies (see Fig. 2d). According to Eq. (1), slopes of linear fits to the $R_{DL}^{2\omega}$ as a function of $1/(H_{ext} - H_K)$ (Fig. 2e) give the information about $H_{DL}$, and the intercepts are the spin Seebeck resistances (or voltages) [21,29].

We calculate $\xi_{DL}$ using $\xi_{DL} = \frac{2eM_s t_{TmIG} H_{DL}}{\hbar J_{ac}}$ [6], where $e$ is the electron charge, $\hbar$ is the reduced Planck constant and $J_{ac}$ is the applied current density. We observe a characteristic increase of $\xi_{DL}$ as $t_{TmIG}$ increases with a saturation length 10 nm (see Fig. 2f). Similarly, previous experiments have revealed a saturation length around 1 nm in ferromagnetic metal heterostructures [18,30,31]. This saturation length is very close to the measured penetration depth of transverse spin current for ferromagnetic metals using spin pumping technique [32-34]. Thus, the saturation length has been interpreted as an indicator of penetration depth [33,34]. By adopting the aforementioned interpretation, our results on the $t_{TmIG}$-dependent $\xi_{DL}$ suggest a penetration depth around 10 nm for TmIG, which is one order of magnitude larger than that of ferromagnetic metals.

**Discussion on $M_S$-dependent SOT efficiency**

Here, we discuss the mechanism for the MI thickness dependence of $\xi_{DL}$. Here, we propose that the $\xi_{DL}$ depends on the $M_S$ when the $M_S$ of thin films is well below the corresponding bulk value. The Landau–Lifshitz–Gilbert equation in the presence of damping-like SOT can be written as

$$M_s t_M \frac{d\hat{m}}{dt} = -\gamma M_s t_M \hat{m} \times \vec{H}_{eff} + \alpha M_s t_M \hat{m} \times \frac{d\hat{m}}{dt} + \gamma J_c \xi_{DL} \frac{\hbar}{2e}(\hat{m} \times \hat{\sigma} \times \hat{m}), \qquad (2)$$

where $\hat{m}$ is the unit vector of magnetization, $\hat{\sigma}$ is the unit vector of current-induced spin polarization, $\gamma$ is the gyromagnetic ratio, $\alpha$ is the Gilbert damping, $t_M$ is the thickness of magnetic layer, $J_c$ is the charge current density, and $\vec{H}_{eff}(=\vec{H}_K + \vec{H}_{ext})$ is the total effective magnetic field acting on the magnetization. The last term on the right-hand side of Eq. 2 arises due to the absorption of transverse spin current by the magnet, which is referred to as the current-induced damping-like (dissipative) SOT. Its strength is parameterized by dimensionless efficiency parameters $\xi_{DL}$. The origin of the SOT can be understood in a simple microscopic picture as follows. A charge current at the heavy metal and ferromagnet interface induces an accumulation of spin density, $\rho\hat{\sigma}$, due to the finite spin-orbit interaction (for example, by SHE or Rashba-Edelstein effect). Here $\rho$ is the magnitude of the spin density, which is proportional to the strength of the spin-orbit interaction. This spin density interacts with the ferromagnet via exchange interaction, of the form $U_{ex} \sim \rho M_s \hat{m} \cdot \hat{\sigma}$, enabling the absorption of the spin current by the ferromagnet. In the perturbative treatment, the spin current absorbed by the ferromagnet can be obtained up to second order in the exchange interaction to yield the damping-like spin-orbit torque with $\xi_{DL} \sim M_S^2$ [35]. The positive correlation between $\xi_{DL}$ and $M_S$ is referred as the $M_S$-effect; it has also been theoretically studied in the frame of spin pumping effect [36], which is the Onsager reciprocal process of the spin torque effect. The increase of spin mixing conductance with $M_S$ is consistent with the calculation from first principles [37] when the surface modification effect presents in the ultrathin regime [38].

Our experiments serve the first demonstration of the $M_S$-effect; we show that as the thickness increases, the SOT efficiency increases with the $M_S$ significantly in the low $M_S$-regime (see Fig. 3), which is in qualitative agreement with the $M_S$-effect. Intuitively, as the magnetic moment density ($M_S$) increases, the interfacial exchange interaction is enhanced, which allows more spin current going through the interface. As the thickness increases, the SOT efficiency saturates earlier than $M_S$, around the half of bulk magnetization (60 emu/cc), which suggests that the SOT is determined by the local magnetization that is saturated at a smaller thickness than the global magnetization $M_S$ [37]. Our experiment is calling for further investigation of the interaction between ultrathin magnetic films and heavy metals, which would include the spin physics of dimensional crossover.

**SOT switching**

After quantifying the SOT efficiency, we perform the CIMS experiments for W/TmIGs with different $t_{TmIG}$. The switching is achieved in all devices with $t_{TmIG}$ up to 15 nm and the switching phase diagrams are summarized in Fig. 4a. In the presence of an external field along the +y direction, a sufficiently large charge current along the +y direction will cause magnetization (AHR) switching from the +z direction to the -z direction (negative to positive). The required amount of charge current to flip the magnetization decreases as the external field increases. When we apply a sufficiently large charge current along the -y direction while keeping the external field along the +y direction, the magnetization (AHR) is switched from the -z direction to the +z direction (positive to negative) (upper panels in Figs. 4b-c). For the same current direction, the switching direction is opposite when we reverse the external field direction (lower panels in Figs. 4b-c). All above facts agree with the picture of SOT-driven magnetization switching. Note that the switching current density is as low as $8 \times 10^{10}$ A/m$^2$ for the W(5 nm)/TmIG (15 nm) (Fig. 4c), which is still two times smaller than the Pt(5 nm)/TmIG(8 nm) case considering that the $t_{TmIG}$ is almost two times larger [15]. This suggests that W enables more energy efficient magnetization switching.

The switching direction driven by current-induced SOTs is consistent with the sign of spin Hall angle of W and is opposite to the Pt/TmIG case [15]. Therefore, our work strongly suggests the dominant role of SHE in the generation of SOTs and CIMS in HM/MI bilayers and complete the paradigm of SOT-driven switching. However, we do notice that there could be an interfacial Rashba-Edelstein effect at W/TmIG interface contributing to the SOTs by comparative analyses of SOTs and SMR(AHR) (see supplementary information).

To quantitatively compare the switching efficiency of W/TmIG devices with different $t_{TmIG}$, we define an effective switching efficiency as $\eta = \frac{2eM_s t_{TmIG} H_P}{\hbar J_{sw}(H_y \to 0)}$ [39], where $H_P$ is the domain wall depinning field estimated from the coercive field and $J_{sw}(H_y \to 0)$ is the zero-field limit of current density in the switching phase diagram. This formula is chosen because the CIMS is achieved through domain nucleation and domain wall motion in the Hall bar devices due to the large scale of our Hall bar devices, of which the channel width is 20 µm. We observe a dramatic increase of $\eta$ with $t_{TmIG}$ (Fig. 4d), for which we consider two reasons. First, the $\xi_{DL}$ increases with $t_{TmIG}$, which means that the same amount of charge current will generate more spin current. Thus, the increase of $\xi_{DL}$ contributes to a lower $J_{sw}$ and thus a larger $\eta$. Second, the Joule heating effect becomes much more significant when a larger charge current is applied, which is the case for switching a thicker TmIG. Joule heating causes reduction of thermal stability through decreasing the $M_S$ and $H_P$; these two values will be smaller than those measured at the low current limit. Therefore, the $M_S$ and $H_P$ used to calculate $\eta$ are overestimated, leading to a larger $\eta$.

**Conclusions**

In summary, we have systematically studied the dimensional crossover of magnetism and its effect on SOTs in ultrathin MI films with perpendicular magnetic anisotropy. The characteristic increase of SOT efficiency with the MI thickness can be understood from the enhancement of magnetic moment density and the suppression of thermal fluctuations. In addition, we have realized CIMS in W/TmIG devices with $t_{TmIG}$ up to 15 nm. The switching current density for W/TmIG devices is lower or comparable with these for HM/ferromagnetic metals despite the fact that the saturated $\xi_{DL}$ is estimated to be only around 0.02 at this stage, which is much less than the 0.3 that is estimated for W in W/CoFeB bilayers [19]. Further improvement of the $\xi_{DL}$ could be done by spin mixing conductance matching [40] and surface treatment [41]. Our results presented here show the great potential of ultrathin MI-based spintronics.

**Methods and Materials**

All TmIG(111) films were grown on $Nd_3Ga_5O_{12}$(111) by pulsed laser deposition [13]. We deposited a 5 nm-thick W layer on top of TmIG followed by subsequent deposition of MgO(2 nm)/$TaO_x$(3 nm) layers to protect W from oxidization. Magnetization hysteresis loops as a function of an out-of-plane magnetic field were measured by a vibrating sample magnetometer and a superconducting quantum interference device. The nominal thin film area is $5 \times 5$ mm². The films were patterned into Hall bar devices (Fig. 2a) by using standard photolithography and dry etching for the resistance, SOT, and switching measurements. The channel width is 20 µm, and the distance between two neighboring Hall contacts is 26 µm. We measured the second harmonic Hall resistance by applying a $I_{ac,r.m.s}$ = 1 mA ($J_{ac,r.m.s}$ = $10^{10}$ A/m²) with a frequency $\omega/2\pi = 195.85$ Hz. The magnetic field and angle controls were done in a physical properties measurement system. The CIMS experiments were performed in the ambient environment by applying a pulse current with 5 ms pulse width and reading Hall voltage subsequently.

**Acknowledgements**

We acknowledge helpful discussions with Sadamichi Maekawa, Chi-Feng Pai, Wei Zhang, Yi Li, Ke Xia, Yongxi Ou and Can Onur Avci. This work is supported as part of the Spins and Heat in Nanoscale Electronic Systems (SHINES), an Energy Frontier Research Center funded by the US Department of Energy (DOE), Office of Science, Basic Energy Sciences (BES), under Award # DE-SC0012670. Q. S., G. Y., A. N., S. A. R., P. U., Q. L. H., L. P., Y. T., and K. W. are also very grateful to the support from the Function Accelerated nanoMaterial Engineering (FAME) Center and Center for Spintronic Materials, Interfaces and Novel Architectures (C-SPIN), two of six centers of Semiconductor Technology Advanced Research network (STARnet), a Semiconductor Research Corporation (SRC) program sponsored by Microelectronics Advanced Research Corporation (MARCO) and Defense Advanced Research Projects Agency (DARPA).

**Author contributions**

Q. S. and G. Y. conceived the project. C. T. and Y. L. grew the thin films. Q. S. and P. Z. performed the SOT measurements. Q. S., G. Y., C. H., and P. Y. performed the resistance measurements. Q. S., P. Z., G. Y. and C. H. performed the switching measurements. A. N. and C. Z. fabricated the Hall bar devices. P. U., S. K., and Y. T. proposed the $M_S$-effect. All authors contributed to the discussion. Q. S., G. Y. and K. L. W. wrote the manuscript with the input from all other authors.

**References**

1. Kajiwara, Y. *et al.* Transmission of electrical signals by spin-wave interconversion in a magnetic insulator. *Nature* **464**, 262-266, doi:10.1038/nature08876 (2010).
2. Bauer, G. E., Saitoh, E. & van Wees, B. J. Spin caloritronics. *Nature materials* **11**, 391-399, doi:10.1038/nmat3301 (2012).
3. Uchida, K. *et al.* Observation of the spin Seebeck effect. *Nature* **455**, 778-781, doi:10.1038/nature07321 (2008).
4. Ming Zhong Wu & Hoffmann, A. *Recent advances in magnetic insulators-from spintronics to microwave applications*. (Acadmeic Press, 2013).
5. Miron, I. M. *et al.* Perpendicular switching of a single ferromagnetic layer induced by in-plane current injection. *Nature* **476**, 189-193, doi:10.1038/nature10309 (2011).
6. Liu, L. *et al.* Spin-torque switching with the giant spin Hall effect of tantalum. *Science* **336**, 555-558, doi:10.1126/science.1218197 (2012).
7. Qiu, X. *et al.* Spin-orbit-torque engineering via oxygen manipulation. *Nature nanotechnology* **10**, 333-338, doi:10.1038/nnano.2015.18 (2015).
8. Yu, G. *et al.* Switching of perpendicular magnetization by spin-orbit torques in the absence of external magnetic fields. *Nature nanotechnology* **9**, 548-554, doi:10.1038/nnano.2014.94 (2014).
9. Liu, L., Lee, O. J., Gudmundsen, T. J., Ralph, D. C. & Buhrman, R. A. Current-Induced Switching of Perpendicularly Magnetized Magnetic Layers Using Spin Torque from the Spin Hall Effect. *Physical review letters* **109**, doi:10.1103/PhysRevLett.109.096602 (2012).
10. Nakayama, H. *et al.* Spin Hall magnetoresistance induced by a nonequilibrium proximity effect. *Physical review letters* **110**, 206601, doi:10.1103/PhysRevLett.110.206601 (2013).
11. Chen, Y.-T. *et al.* Theory of spin Hall magnetoresistance. *Physical Review B* **87**, doi:10.1103/PhysRevB.87.144411 (2013).
12. Hahn, C. *et al.* Comparative measurements of inverse spin Hall effects and magnetoresistance in YIG/Pt and YIG/Ta. *Physical Review B* **87**, doi:10.1103/PhysRevB.87.174417 (2013).
13. Tang, C. *et al.* Anomalous Hall hysteresis in $Tm_3Fe_5O_{12}$/Pt with strain-induced perpendicular magnetic anisotropy. *Physical Review B* **94**, doi:10.1103/PhysRevB.94.140403 (2016).
14. Li, P. *et al.* Spin-orbit torque-assisted switching in magnetic insulator thin films with perpendicular magnetic anisotropy. *Nat Commun* **7**, 12688, doi:10.1038/ncomms12688 (2016).
15. Avci, C. O. *et al.* Current-induced switching in a magnetic insulator. *Nature materials* **16**, 309-314, doi:10.1038/nmat4812 (2017).
16. Li, J. *et al.* Deficiency of the bulk spin Hall effect model for spin-orbit torques in magnetic-insulator/heavy-metal heterostructures. *Physical Review B* **95**, doi:10.1103/PhysRevB.95.241305 (2017).
17. Zhang, W., Han, W., Jiang, X., Yang, S.-H. & S. P. Parkin, S. Role of transparency of platinum–ferromagnet interfaces in determining the intrinsic magnitude of the spin Hall effect. *Nature Physics* **11**, 496-502, doi:10.1038/nphys3304 (2015).
18. Pai, C.-F., Ou, Y., Vilela-Leão, L. H., Ralph, D. C. & Buhrman, R. A. Dependence of the efficiency of spin Hall torque on the transparency of Pt/ferromagnetic layer interfaces. *Physical Review B* **92**, doi:10.1103/PhysRevB.92.064426 (2015).
19. Pai, C.-F. *et al.* Spin transfer torque devices utilizing the giant spin Hall effect of tungsten. *Appl Phys Lett* **101**, 122404, doi:10.1063/1.4753947 (2012).
20. Garello, K. *et al.* Symmetry and magnitude of spin-orbit torques in ferromagnetic heterostructures. *Nature nanotechnology* **8**, 587-593, doi:10.1038/nnano.2013.145 (2013).
21. Avci, C. O. *et al.* Interplay of spin-orbit torque and thermoelectric effects in ferromagnet/normal-metal bilayers. *Physical Review B* **90**, doi:10.1103/PhysRevB.90.224427 (2014).
22. Paoletti, A. *Physics of Magnetic Garnets*. (North-Holland Publishing Company, 1978).
23. Tang, C. *et al.* Above 400-K robust perpendicular ferromagnetic phase in a topological insulator. *Science advances* **3**, e1700307, doi:10.1126/sciadv.1700307 (2017).
24. Zhang, R. & Willis, R. F. Thickness-dependent Curie temperatures of ultrathin magnetic films: effect of the range of spin-spin interactions. *Physical review letters* **86**, 2665-2668, doi:10.1103/PhysRevLett.86.2665 (2001).
25. Huang, F., Kief, M. T., Mankey, G. J. & Willis, R. F. Magnetism in the few-monolayers limit: A surface magneto-optic Kerr-effect study of the magnetic behavior of ultrathin films of Co, Ni, and Co-Ni alloys on Cu(100) and Cu(111). *Physical Review B* **49**, 3962-3971, doi:10.1103/PhysRevB.49.3962 (1994).
26. Vaz, C. A. F., Bland, J. A. C. & Lauhoff, G. Magnetism in ultrathin film structures. *Reports on Progress in Physics* **71**, 056501, doi:10.1088/0034-4885/71/5/056501 (2008).

| | |
|---|---|
| 27 | Li, Y. & Baberschke, K. Dimensional crossover in ultrathin Ni(111) films on W(110). *Physical review letters* **68**, 1208-1211, doi:10.1103/PhysRevLett.68.1208 (1992). |
| 28 | Shao, Q. *et al.* Strong Rashba-Edelstein Effect-Induced Spin-Orbit Torques in Monolayer Transition Metal Dichalcogenide/Ferromagnet Bilayers. *Nano letters* **16**, 7514-7520, doi:10.1021/acs.nanolett.6b03300 (2016). |
| 29 | Uchida, K.-i. *et al.* Observation of longitudinal spin-Seebeck effect in magnetic insulators. *Appl Phys Lett* **97**, 172505, doi:10.1063/1.3507386 (2010). |
| 30 | Kim, J. *et al.* Layer thickness dependence of the current-induced effective field vector in Ta|CoFeB|MgO. *Nature materials* **12**, 240-245, doi:10.1038/nmat3522 (2013). |
| 31 | Pai, C.-F., Mann, M., Tan, A. J. & Beach, G. S. D. Determination of spin torque efficiencies in heterostructures with perpendicular magnetic anisotropy. *Physical Review B* **93**, doi:10.1103/PhysRevB.93.144409 (2016). |
| 32 | Ghosh, A., Auffret, S., Ebels, U. & Bailey, W. E. Penetration depth of transverse spin current in ultrathin ferromagnets. *Physical review letters* **109**, 127202, doi:10.1103/PhysRevLett.109.127202 (2012). |
| 33 | Qiu, X. *et al.* Enhanced Spin-Orbit Torque via Modulation of Spin Current Absorption. *Physical review letters* **117**, 217206, doi:10.1103/PhysRevLett.117.217206 (2016). |
| 34 | Ou, Y., Pai, C.-F., Shi, S., Ralph, D. C. & Buhrman, R. A. Origin of fieldlike spin-orbit torques in heavy metal/ferromagnet/oxide thin film heterostructures. *Physical Review B* **94**, doi:10.1103/PhysRevB.94.140414 (2016). |
| 35 | Bender, S. A. & Tserkovnyak, Y. Interfacial spin and heat transfer between metals and magnetic insulators. *Physical Review B* **91**, doi:10.1103/PhysRevB.91.140402 (2015). |
| 36 | Ohnuma, Y., Adachi, H., Saitoh, E. & Maekawa, S. Enhanced dc spin pumping into a fluctuating ferromagnet nearTC. *Physical Review B* **89**, doi:10.1103/PhysRevB.89.174417 (2014). |
| 37 | Jia, X., Liu, K., Xia, K. & Bauer, G. E. W. Spin transfer torque on magnetic insulators. *EPL (Europhysics Letters)* **96**, 17005, doi:10.1209/0295-5075/96/17005 (2011). |
| 38 | Farle, M. Ferromagnetic resonance of ultrathin metallic layers. *Reports on Progress in Physics* **61**, 755-826, doi:10.1088/0034-4885/61/7/001 (1998). |
| 39 | Lee, O. J. *et al.* Central role of domain wall depinning for perpendicular magnetization switching driven by spin torque from the spin Hall effect. *Physical Review B* **89**, doi:10.1103/PhysRevB.89.024418 (2014). |
| 40 | Du, C., Wang, H., Yang, F. & Hammel, P. C. Enhancement of Pure Spin Currents in Spin PumpingY3Fe5O12/Cu/MetalTrilayers through Spin Conductance Matching. *Physical Review Applied* **1**, doi:10.1103/PhysRevApplied.1.044004 (2014). |
| 41 | Jungfleisch, M. B., Lauer, V., Neb, R., Chumak, A. V. & Hillebrands, B. Improvement of the yttrium iron garnet/platinum interface for spin pumping-based applications. *Appl Phys Lett* **103**, 022411, doi:10.1063/1.4813315 (2013). |

**Figures and legends**

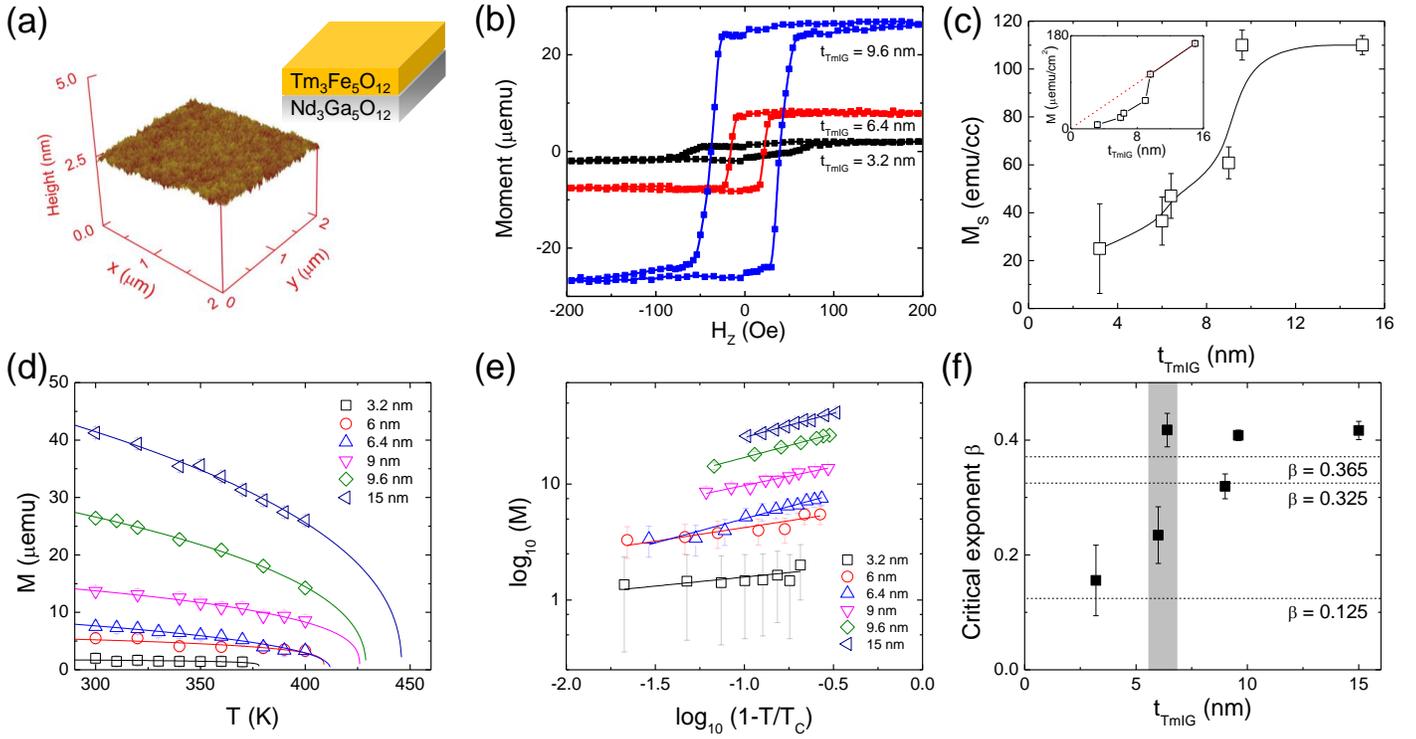

**Figure 1.** Dimensional crossover of magnetism in TmIG thin films. (a) Atomic force microscopy image of a 10 nm-thick TmIG film. (b) Magnetic moment as a function of out-of-plane magnetic field for TmIG thin films with different thicknesses at room temperature. (c) Saturation magnetization as a function of TmIG thickness at room temperature. The inset shows the areal magnetic moment as a function of TmIG thickness, which indicates a negligible magnetic dead layer. (d) Total magnetic moment as a function of temperature for different TmIG thicknesses. The solid lines are power-law fits to $M = M_0 \left(1 - T/T_C\right)^\beta$. (e) $\log_{10}(M)$ vs $\log_{10}(1-T/T_C)$ plots from (d) showing the thickness dependence of the $\beta$ values. (f) Critical exponent $\beta$ vs TmIG thickness showing a dimensional crossover from 2D to 3D. The dashed lines are theoretical values for 2D Ising ($\beta = 0.125$), 3D Ising ($\beta = 0.325$) and 3D Heisenberg ($\beta = 0.365$) models.

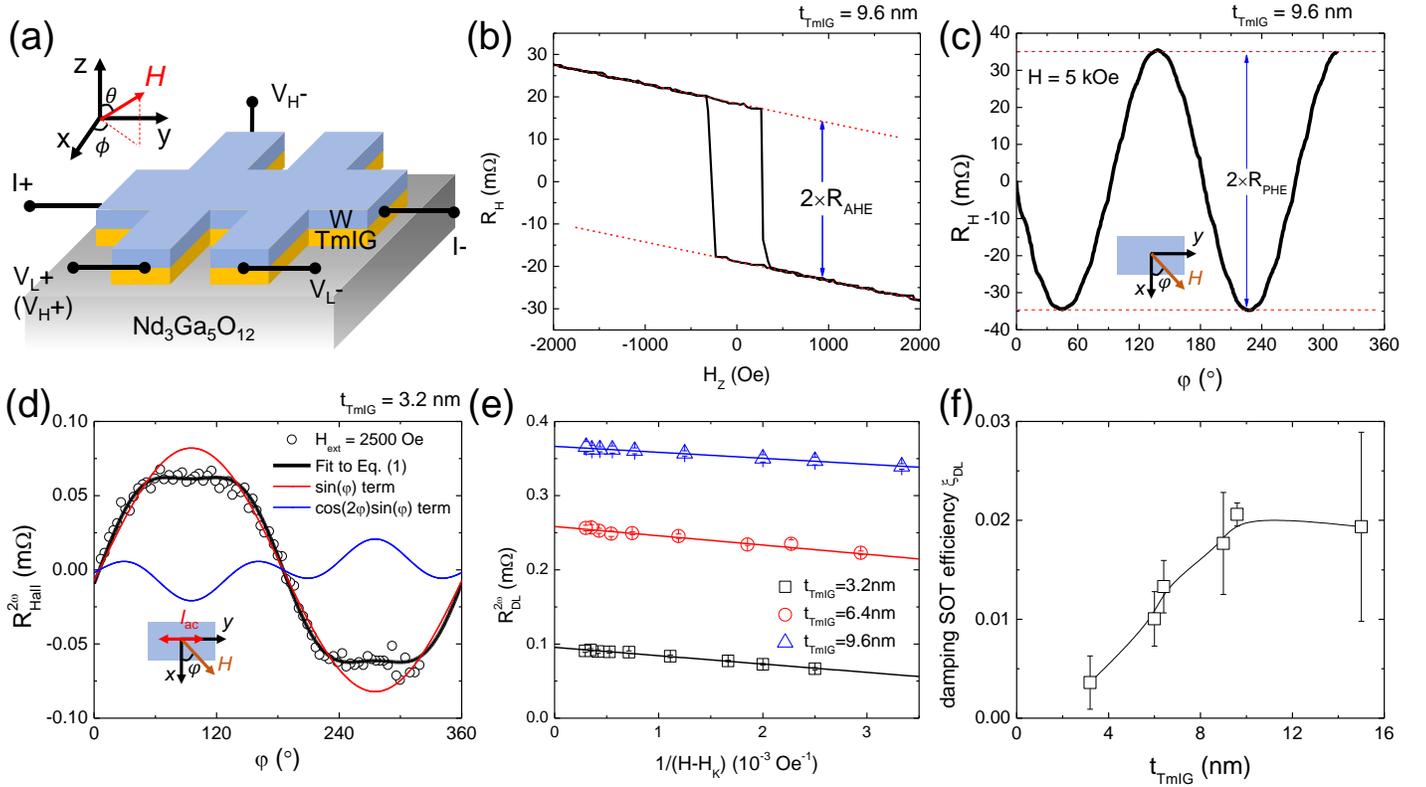

**Figure 2.** Spin transport and SOT measurements in the W/TmIG bilayers. (a) Experimental setup for measuring resistance, SOT and current-induced magnetization switching. (b) Hall resistance as a function of an out-of-plane magnetic field for the W (5 nm)/TmIG (9.6 nm), where AHE is observed as the sharp square hysteresis loop. (c) Hall resistance as a function of a rotating in-plane constant magnetic field (5 kOe) for the W (5 nm)/TmIG(9.6 nm), where SMR-induced PHE is observed. (d) Second-harmonic Hall resistance as a function of in-plane azimuthal angle for the external magnetic field 2500 Oe for the W (5 nm)/TmIG (3.2 nm), where the black curve is the fit to Eq. (1). Both $\cos 2\varphi \sin \varphi$ (blue curve) and $\sin \varphi$ (red curve) angle dependencies are revealed. (e) Extracted damping-like torque contribution as a function of the inverse of external magnetic field subtracting the anisotropy field. The large intercepts are the spin Seebeck resistance. (f) Damping-like spin-orbit torque efficiency as a function of TmIG thickness.

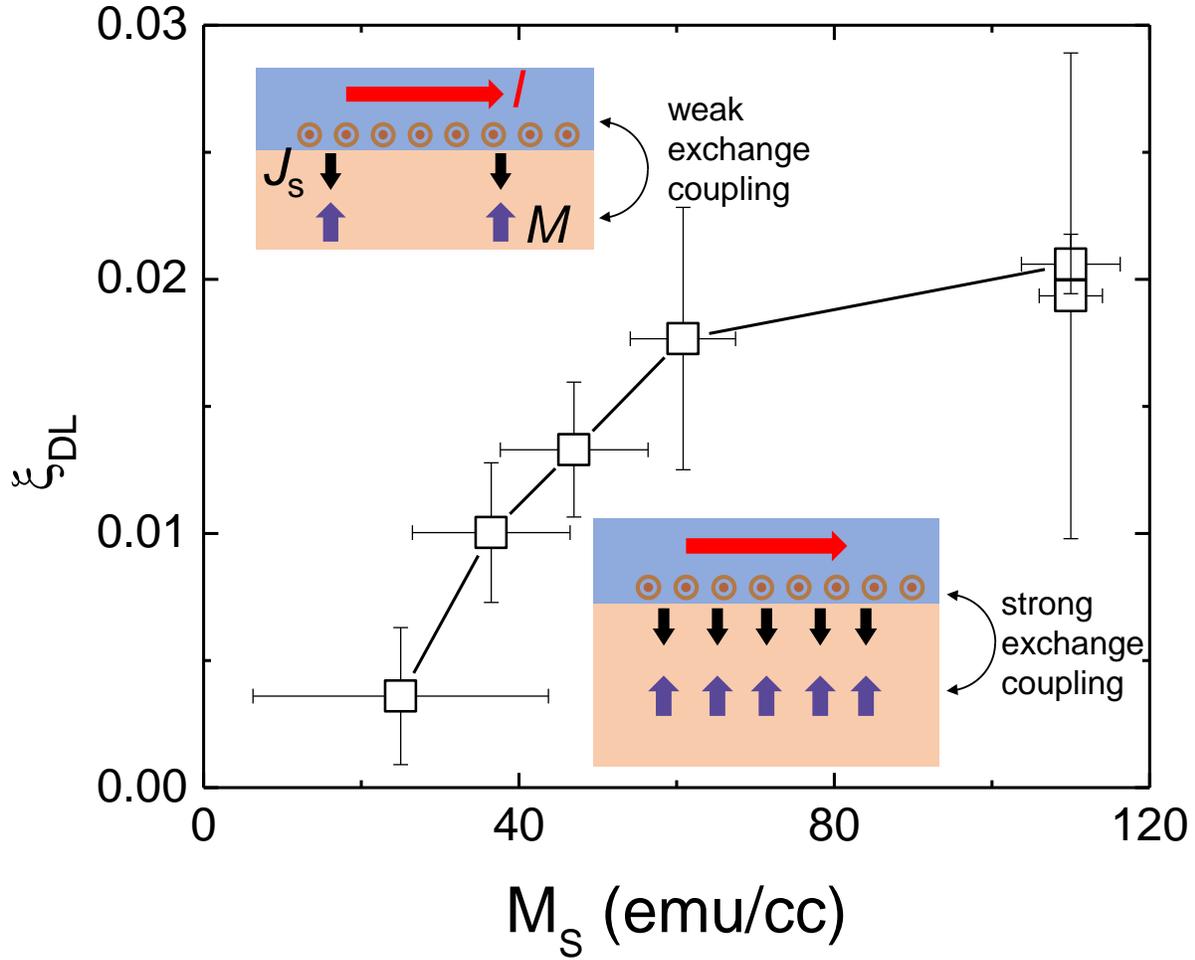

**Figure 3.** Role of TmIG $M_S$ on the $\xi_{DL}$. $\xi_{DL}$ is proportional to the $M_S$ squared as shown in the text when the $M_S$ is small due to strong thermal fluctuation and surface modification effect. Insets show two cases: in the left inset, the magnetic moment density is small and thus the interfacial exchange interaction is weak, resulting in a small spin current injection; in the right inset, the magnetic moment density is large due to suppressed thermal fluctuation and thus the interfacial exchange interaction is strong, resulting in a large spin current injection.

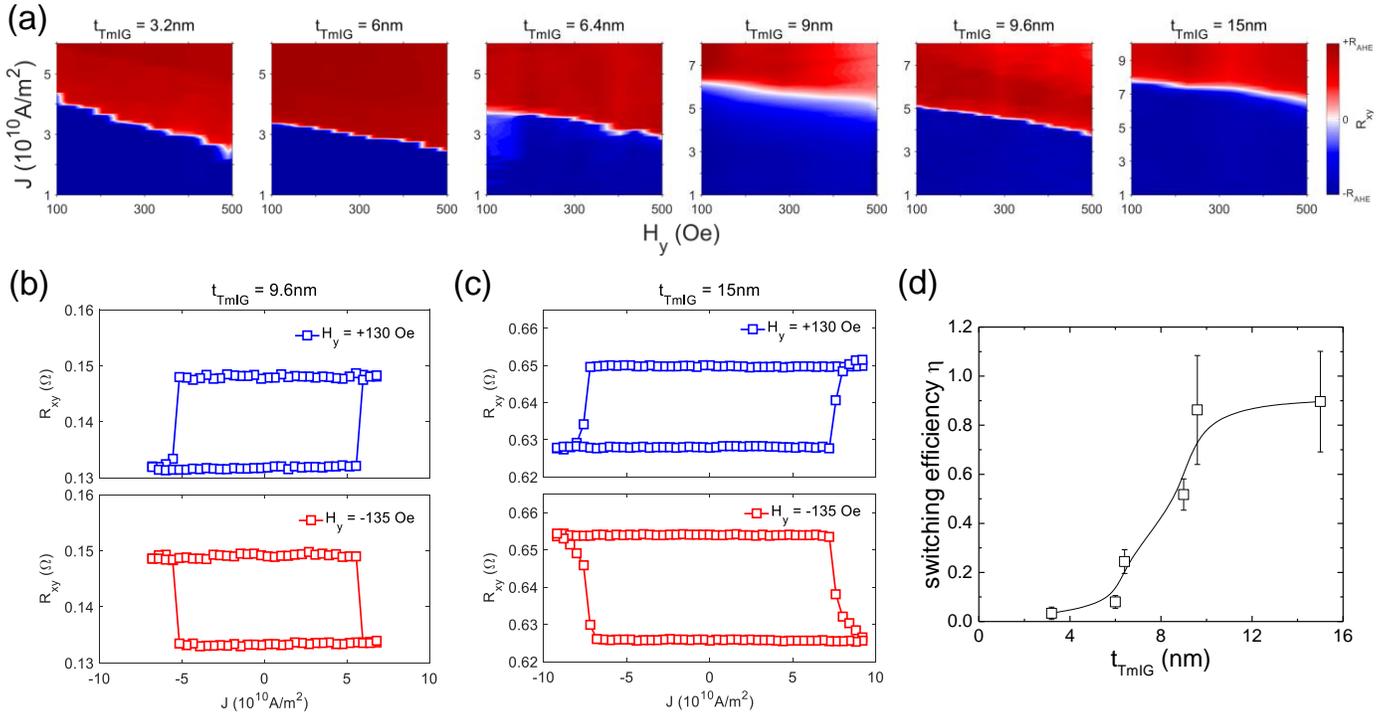

**Figure 4.** Current-induced magnetization switching in W/TmIG with different TmIG thicknesses. (a) Switching phase diagram for TmIG with thicknesses from 3.2 nm to 15 nm, where the external field is along the current direction. For instances, (b) and (c) show the current-induced switching for TmIG with thickness 9.6 nm and 15 nm, respectively, in the presence of a magnetic field along and against the current direction. The switching is done by applying a 5 ms pulse with varying current amplitude. (d) TmIG thickness dependent current switching efficiency, which is estimated from the depinning (coercive) field over switching current density in the zero-external field limit.
17